\documentclass{PoS}
\usepackage{amssymb,amsmath}
\usepackage{tabularx}
\def\beq{\begin{equation}}
\def\eeq{\end{equation}}
\def\bea{\begin{eqnarray}}
\def\eea{\end{eqnarray}}

\title{Polarisabilities of the nucleon in baryon chiral perturbation theory and beyond}

\ShortTitle{Polarisabilities of the nucleon in $B\chi$PT and beyond}

\author{\speaker{Vadim Lensky}\\
        Johannes-Gutenberg Universit\"at Mainz, D-55099 Mainz, Germany\\
        E-mail: \email{vlenskiy@uni-mainz.de}}

\author{Vladimir Pascalutsa\\
        Johannes-Gutenberg Universit\"at Mainz, D-55099 Mainz, Germany\\
        E-mail: \email{vladipas@uni-mainz.de}}

\abstract{We review the recent baryon chiral perturbation theory results for the
nucleon polarisabilities that describe the different regimes of nucleon Compton scattering --- real,
virtual, and doubly virtual. We stress the importance of the empirical verification of the theory
in the context of the calculation of the inelastic nucleon structure corrections, such as the two-photon exchange contributions.
We also discuss the recently obtained constraints that relate the different
regimes of nucleon Compton scattering and can provide additional information on the nucleon structure.}

\FullConference{The 9th International workshop on Chiral Dynamics\\
		17-21 September 2018\\
		Durham, NC, USA}

\begin{document}

\section{Introduction}
The electromagnetic structure of the nucleon in the form of charge/magnetisation distribution
and inelastic structure functions has received a new round of revision recently in 
connection with the ``proton radius puzzle''. While the puzzle concerns the proton charge
radius (or, more generally, the electric form factor at low $Q$), the inelastic
structure, parametrised by polarisabilities, enters prominently in the calculation
of subleading corrections, such as the two-photon exchange. These calculations require
a comprehensive understanding of the nucleon Compton scattering (CS) with both photons virtual.
The experimentally accessible CS regimes are, however, at present limited to
\begin{itemize}
\item  real CS (RCS), with both photons real: $q^2 = q^{\prime 2} = 0$,
\item virtual CS (VCS), with the initial photon having a non-zero virtuality $q^2 = -Q^2 <0$ and the final one being real, $q^{\prime 2}=0$,
\item forward doubly virtual CS (VVCS), where both photons are virtual and have the same momentum, $q = q'$, $q^2 = -Q^2 <0$,
\end{itemize}
where the information on the latter process is available from the moments of the nucleon structure functions;
see the reviews~\cite{Guichon:1998xv,Drechsel:2002ar,Schumacher:2005an,Holstein:2013kia,Hagelstein:2015egb,Griesshammer:2012we}
for detailed information on nucleon CS in these regimes.

Another source of information on the nucleon CS follows from the unitarity and analyticity of the CS amplitude, which allows one to derive model-independent sum rules for various electromagnetic
structure quantities~\cite{GellMann:1954db}.
These sum rules relate a low-energy quantity with a weighted integral
of a photoabsorption cross section on the nucleon, some of the well-known examples being the Baldin sum rule for the sum of the dipole (static) polarisabilities~\cite{Baldin},
the Gerasimov-Drell-Hearn (GDH) sum rule for the anomalous magnetic moment~\cite{Gerasimov:1965et,Drell:1966jv}, both derived for RCS,
and the Burkhardt-Cottingham sum rule~\cite{Burkhardt:1970ti}, derived for VVCS. 
The analyticity of the most general (non-Born) CS amplitude provides constraints between the polarisabilities parametrising the three different regimes --- RCS, VCS, and VVCS --- 
due to these regimes being special cases of the most general CS kinematics.

A natural way to analyse the wealth of experimental data on all CS regimes
is to use the chiral perturbation theory ($\chi$PT) 
\cite{Weinberg:1978kz,Gasser:1983yg,GSS89}. As a low-energy
effective-field theory of QCD, it provides the model-independent link between the nucleon Compton-scattering data and polarisabilities. 
It should also naturally yield CS amplitudes that satisfy the analyticity constraints.
Using $\chi$PT, one can perform calculations in the different regimes of nucleon CS in the same unified framework and test the theory against the data. This verification allows one to reliably
calculate the two-photon exchange corrections starting from the $\chi$PT result
for the nucleon CS amplitude. Here we review the recent results of this approach.


\section{Nucleon Compton scattering in covariant baryon $\chi$PT}

We consider nucleon CS in baryon $\chi$PT (B$\chi$PT), a manifestly-covariant formulation of 
$\chi$PT with pion, nucleon and $\Delta(1232)$ isobar degrees of freedom, see ref.~\cite{Pascalutsa:2006up} for review.
The details of the calculation
can be found in~\cite{Lensky:2008re,Lensky:2009uv,Lensky:2014efa,Lensky:2015awa} for RCS, \cite{Lensky:2016nui} for VCS, and \cite{Lensky:2014dda} for VVCS. 

Our effective field theory expansion uses the $\delta$-counting~\cite{Pascalutsa:2002pi}: the mass difference between the nucleon and the $\Delta$ isobar, $\varDelta=M_\Delta-M$, is considered
an intermediate scale, so that
$m_\pi/\varDelta\simeq\varDelta/\Lambda_\chi \equiv\delta$,
and hence the usual chiral expansion scale $m_\pi/\Lambda_\chi$ is counted as $\delta^2$.
This expansion scheme, in particular, naturally generates the Delta resonance width necessary
for the description of the resonance peak.

\begin{figure}[h]\centering
\includegraphics[width=.4\columnwidth]{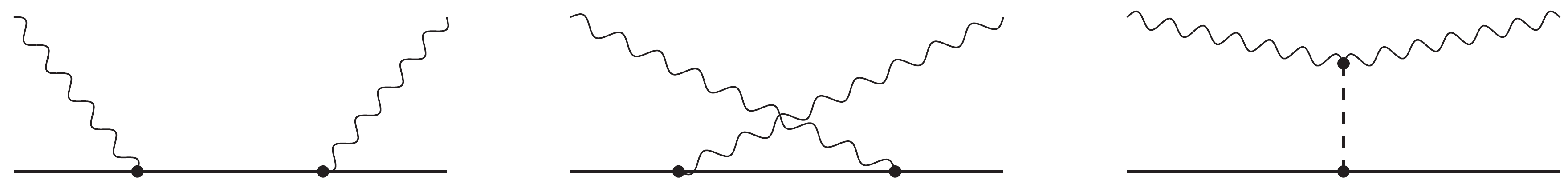}
\caption{Born graphs and the anomaly graph.}
\label{fig:Born}
\end{figure}

\begin{figure}[h]
\begin{minipage}{0.4\textwidth}
\centering
       \includegraphics[width=\textwidth]{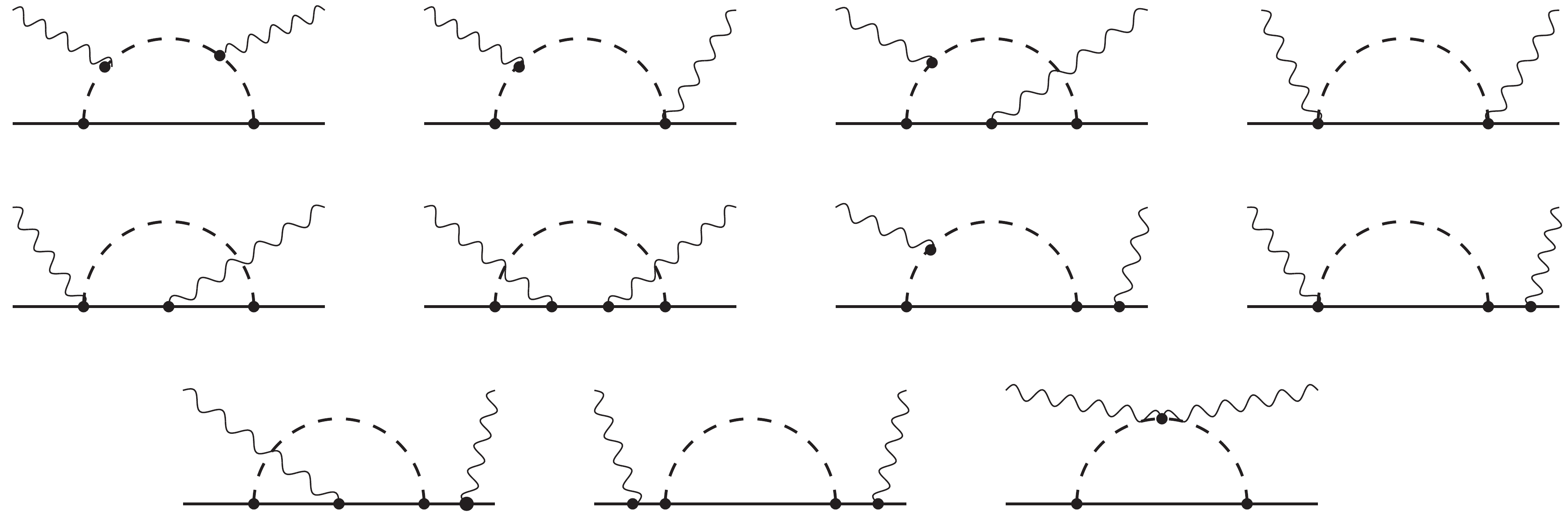}
    \end{minipage}\hfill
\begin{minipage}{0.4\textwidth}
\centering
  \includegraphics[width=\textwidth]{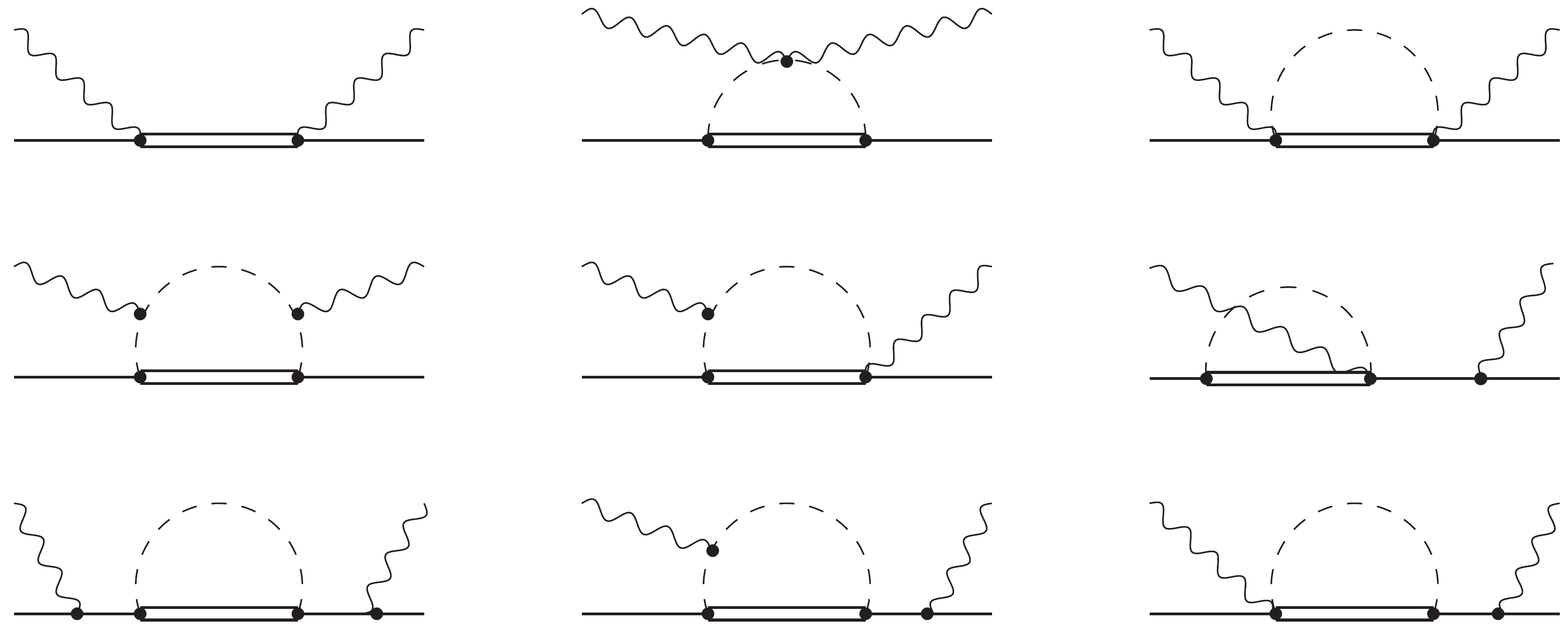}
\end{minipage}
\caption{Left: $\pi N$ loops that contribute to nucleon polarisabilities at NLO; Right: $\pi \Delta$ loops and the $\Delta$ pole graph that contribute to nucleon polarisabilities at NNLO.
 Crossed and time-reversed graphs are not shown but are included in the calculation.}
\label{fig:Loops}
\end{figure}

The graphs that enter our B$\chi$PT calculation are shown in figs.~\ref{fig:Born} and~\ref{fig:Loops}. At leading order (LO), there are the Born graphs and the pion anomaly graph; the former
do not contribute to the polarisabilities, while the latter gives a large and well-known contribution to the spin polarisabilities and is customarily
separated and treated together with the Born graphs. The leading contribution to the polarisabilities (with the pion anomaly contribution subtracted) comes 
from the $\pi N$ loops in fig.~\ref{fig:Loops} at next-to-leading order (NLO), whereas the $\pi \Delta$ loops and the $\Delta$ pole graph in the same figure contribute at next-to-next-to-leading order (NNLO).
The NNLO B$\chi$PT calculation does not involve any unknown parameters (they start to appear at one order higher).
Note that in the VCS and the VVCS cases the photon-nucleon vertex and the photon-nucleon-$\Delta$ vertex acquire form factors that depend on the photon virtuality;
see the details in the respective references cited above.

\section{Nucleon RCS and polarisabilities in B$\chi$PT}

\begin{figure}[h]
\centering
\begin{minipage}{0.8\textwidth}
\centering
       \includegraphics[width=\textwidth]{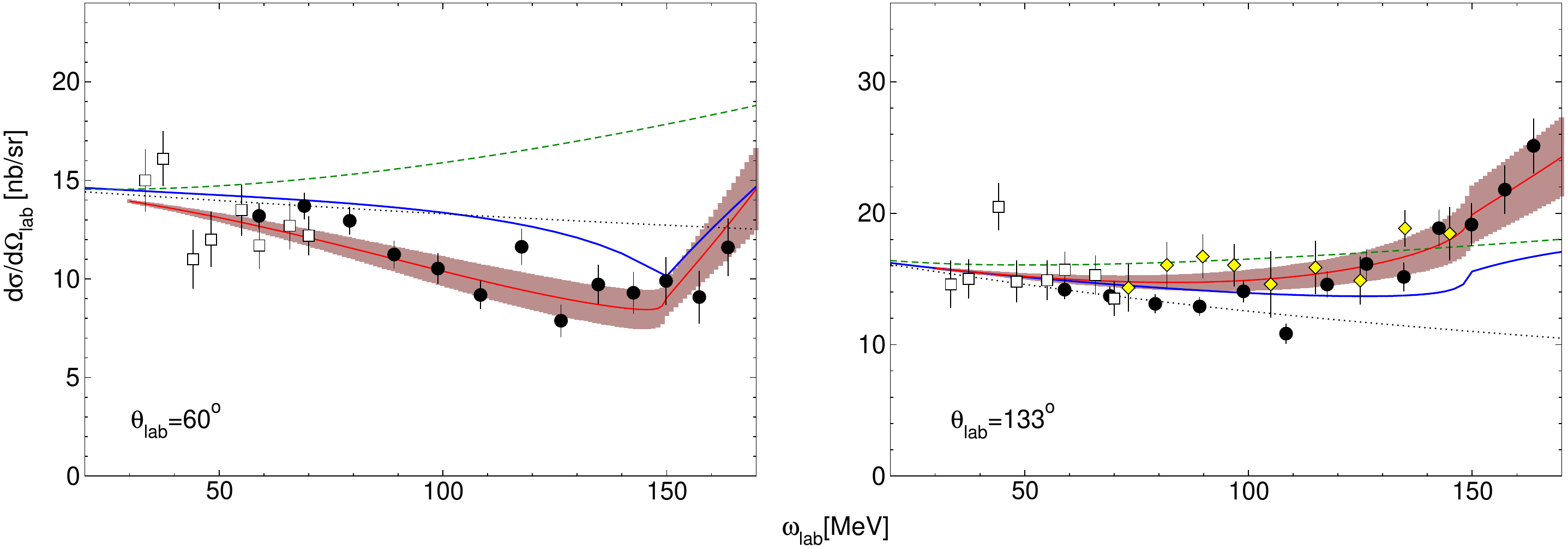}
    \end{minipage}
\caption{The result of the B$\chi$PT calculation for the proton RCS cross section compared with experimental data. Data points are from: Illinois~\cite{Federspiel:1991yd} --- open squares, SAL~\cite{MacG95} ---
open diamonds, and MAMI~\cite{MAMI01} --- filled circles. The curves are: black dotted --- Born graphs only, green dashed --- Born$+$anomaly, blue solid --- Born$+$anomaly$+\pi N$ loops, 
red solid with a band --- full NNLO B$\chi$PT.}
\label{fig:chipt_xs}
\end{figure}

The definition of the RCS static polarisabilities is connected to the expansion of the RCS amplitude in powers of the photon energy $\omega$,
which is parametrised, up to terms of $O(\omega^4)$, by the following polarisabilities~\cite{Babusci:1998ww,Holstein:1999uu}:
%
%
$\alpha_{E1}$ and $\beta_{M1}$ --- the static electric and magnetic dipole polarisabilities at $O(\omega^2)$; four static spin polarisabilities $\gamma_{E1E1},\ \gamma_{M1M1},\ \gamma_{E1M2},\ \text{and }\gamma_{M1E2}$ at $O(\omega^3)$,
and the electric and magnetic dispersive and quadrupole polarisabilities, $\alpha_{E1\nu}$, $\beta_{M1\nu}$, $\alpha_{E2}$, and $\beta_{M2}$, enter at $O(\omega^4)$.
%
%
Figure~\ref{fig:chipt_xs} and tables~\ref{tab:polsscalarproton} and~\ref{tab:polsspinproton}
show how these quantities, along with the experimental data on the unpolarised cross section, are reproduced in B$\chi$PT.
While the $\pi N$ loops are necessary in order to describe the pion production threshold behaviour,
the $\Delta$ pole and the $\pi \Delta$ loops 
bring the polarisabilities close to their physical values, resulting in a good description
of the available low-energy experimental data at NNLO, as illustrated in fig.~\ref{fig:chipt_xs}.
This is further explained in table~\ref{tab:polsscalarproton}; one can see that the $\Delta$ pole contribution,
in particular, to $\beta_{M1}$, is quite large and is nicely accommodated to provide the final value.
B$\chi$PT is in a reasonably good agreement with the PDG values of $\alpha_{E1}$ and $\beta_{M1}$, with the latter being somewhat larger in B$\chi$PT. This seems to be a feature of all $\chi$PT
extractions as contrasted to the dispersion relations (DR) results
that tend to obtain a smaller value of the magnetic dipole polarisability, see~\cite{Lensky:2015awa}
for further comparison.
It has been suggested recently in a partial wave analysis of proton RCS data below
the pion production threshold~\cite{Krupina:2017pgr} that this 
difference might be a problem of the experimental database,
see also refs.~\cite{Pasquini:2017ehj,Pasquini:2019nnx}. This could be further tested when the new RCS data become available~\cite{Downie_talk,Martel_talk,Miskimen_talk}.
\def\hpm#1{\hphantom{#1}}
\begin{table}[ht]
\begin{tabular}{c|c|c|c|c|c|c}
  Source      &$\alpha_{E1}$    & $\beta_{M1}$     &$\alpha_{E2}$  &$\beta_{M2}$ &$\alpha_{E1\nu}$ &  $\beta_{M1\nu}$\\
\hline
$\pi N$ loops & $\hpm{-}6.9 $   &      $-1.8$     & $\hpm{-}13.5$       & $-8.4$  & $\hpm{-}0.7$ & $1.8$   \\
$\pi\Delta$
        loops & $\hpm{-}4.4 $   &    $-1.4$       & $\hpm{-}3.2$        & $-2.7$  & $-0.6$ & $0.6$     \\
$\Delta$ pole &   $-0.1 $       & $\hpm{-}7.1$    & $\hpm{-}0.6$        & $-4.5$      & $-1.5$ & $4.7$ \\
Total         & $ 11.2\pm0.7 $  &$\hpm{-}3.9\pm0.7$& $\hpm{-}17.3\pm3.9$       & $-15.5\pm3.5$    & $-1.3\pm 1.0$ & $7.1\pm 2.5$   \\
\hline
\hline
PDG~\cite{Agashe:2014kda}
             & $11.2\pm 0.4$   & $2.5\pm 0.4$     &     $\cdots $    &  $\cdots $     & $\cdots $   & $\cdots $  \\
\end{tabular}
\caption{Values of proton static scalar polarisabilities, in units of $10^{-4}$~fm$^{3}$ (dipole) and $10^{-4}$~fm$^{5}$ (quadrupole and dispersive).
}
\label{tab:polsscalarproton}
\end{table}
\begin{table}[ht]
\begin{center}
\begin{tabular}{c|c|c|c|c}
  Source           & $\gamma_{E1E1}$ & $\gamma_{M1M1}$& $\gamma_{E1M2}$& $\gamma_{M1E2}$ \\ 
\hline
$\pi N$ loops      &   $-3.4 $      &     $-0.1$    &     $\hpm{-} 0.5  $  &   $\hpm{-} 0.9  $ \\
$\pi\Delta$ loops  &    $\hpm{-}0.4 $      &     $-0.2$    &     $\hpm{-} 0.1  $  &   $-0.2  $  \\
$\Delta$ pole      &   $-0.4 $      &     $\hpm{-}3.3 $    &     $-0.4  $  &   $\hpm{-} 0.4  $   \\
Total              &$-3.3\pm 0.8$   &$2.9\pm 1.5$   &$ 0.2\pm 0.2 $ & $1.1 \pm0.3 $  \\ 
\hline
\hline
MAMI 2015~\cite{Martel:2014pba}
                    &  $-3.5\pm 1.2$  &$3.16\pm 0.85$  & $-0.7\pm 1.2$  & $1.99\pm 0.29$ \\ 
\end{tabular}
\caption{Values of proton static spin polarisabilities, in units of $10^{-4}$~fm$^{4}$.
}
\label{tab:polsspinproton}
\end{center}
\end{table}
B$\chi$PT also agrees quite well with the empirically extracted spin polarisabilities~\cite{Martel:2014pba}. 
The $\chi$PT description of the low-energy RCS data does not seem to be very sensitive to the values of the lowest spin polarisabilities.


\section{Nucleon VCS and generalised polarisabilities in B$\chi$PT}

\begin{figure}[h]
\centering
\begin{minipage}{0.8\textwidth}
\centering
       \includegraphics[width=\textwidth]{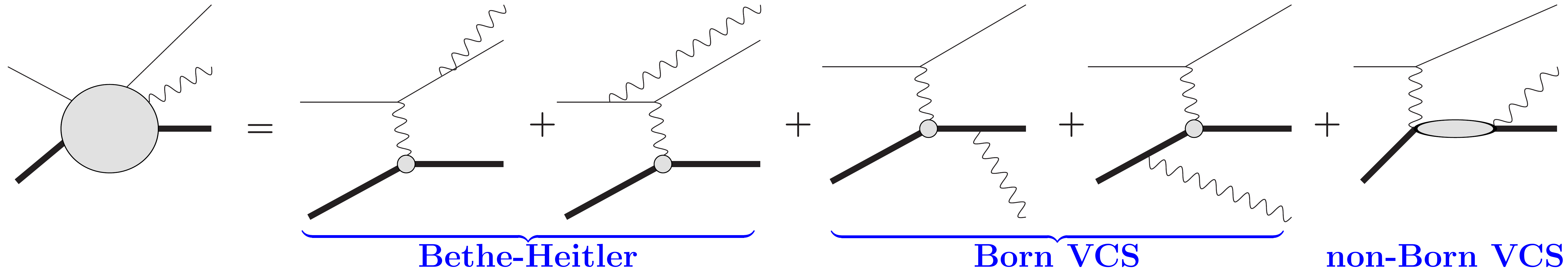}
    \end{minipage}
\caption{The mechanisms that contribute to the process $ep\to ep\gamma$. The contribution of the last graph is parametrised by the generalised polarisabilities (GPs).}
\label{fig:VCS_VCS}
\end{figure}

Nucleon VCS is experimentally accessible in the process $ep\to ep\gamma$, and in the conventional setup the energy $\omega'$ of the final photon is considered small
so that one can expand the amplitude and the observables in powers of $\omega'$. The (spacelike) virtuality of the initial photon $q^2 = -Q^2$
can be arbitrary but not too large so as the B$\chi$PT treatment can still be applied to VCS.
The two leading terms $\propto \omega^{\prime -1}$ and $\propto \omega^{\prime 0}$ 
in the expansion in powers of $\omega'$ are given by the Bethe-Heitler (BH) and Born terms,
 see fig.~\ref{fig:VCS_VCS},
with the generalised polarisabilities (GPs) starting to contribute at $O(\omega')$.
The GPs are functions of $Q^2$ that
are simply linear combinations of the VCS amplitudes taken at the special kinematics $\omega'=0$, $t = -Q^2$ [which also means $\omega = Q^2/(2M)$], see~\cite{Guichon:1995pu,Drechsel:1997xv,Drechsel:1998zm,Gorchtein:2009qq,Lensky:2017dlc} for the definitions. 
At the order $O(\omega')$, there are six GPs, conventionally denoted as 
\beq
P^{(L1,L1)0}(Q^2),\ P^{(M1,M1)0}(Q^2),\ P^{(L1,L1)1}(Q^2),\ P^{(M1,M1)1}(Q^2),\
P^{(M1,L2)1}(Q^2),\ P^{(L1,M2)1}(Q^2)\,,
\label{eq:GPs}
\eeq
where $L1$ or $M2$ etc.\ in the superscript denotes whether the photon is of the longitudinal/electric or the magnetic type, with the corresponding angular momentum
(for the final or initial photon, in that order), whereas the last number in the superscript indicates whether the transition involves the proton's spin flip $(1)$ or not $(0)$.

The natural generalisations
of the static polarisabilities to the VCS case are
\begin{align}
\alpha_{E1}(Q^2)&=-\mbox{$\frac{e^2}{4\pi}\sqrt{\frac{3}{2}}$}P^{(L1,L1)0}(Q^2)\,,\qquad &\beta_{M1}(Q^2)&=-\mbox{$\frac{e^2}{4\pi}\sqrt{\frac{3}{8}}$} P^{(M1,M1)0}(Q^2)\,,\nonumber\\
\gamma_{M1E2}(Q^2) & = -\mbox{$\frac{e^2}{4\pi}\frac{3}{2}\sqrt{\frac{3}{2}}$}P^{(L1,L1)0}(Q^2)\,, \qquad &\gamma_{E1M2}(Q^2) &= -\mbox{$\frac{e^2}{4\pi}\frac{3}{\sqrt{2}}$} P^{(M1,M1)0}(Q^2),
\end{align}
where $e$ is the proton charge. These functions coincide with the corresponding static RCS polarisabilities at $Q^2=0$~\cite{Guichon:1995pu,Drechsel:1998zm}.
These definitions are specific for VCS, and should be distinguished
from the corresponding definitions for VVCS (which are made at a different kinematics).
The remaining two GPs, $P^{(L1,L1)1}$ and $P^{(M1,M1)1}$, vanish at $Q^2=0$. Their
slopes are related to other static polarisabilities and to the VVCS GPs via the
spin-dependent sum rules.

The unpolarised differential cross-section of the process $ep\to ep\gamma$, up to terms linear in $\omega'$, depends only on the three VCS response functions,
see, e.g., ref.~\cite{Guichon:1998xv,Guichon:1995pu}:
\begin{align}
P_{LL}(Q^2)&=-2\sqrt{6}M G_E(Q^2) P^{(L1,L1)0}(Q^2)\,,\nonumber\\
P_{TT}(Q^2)&=\hphantom{-}6 M G_M(Q^2)(1+\tau)
\left[
2\sqrt{2}\,M \tau\, P^{(L1,M2)1}(Q^2)+P^{(M1,M1)1}(Q^2)
\right]
\,,\nonumber\\
P_{LT}(Q^2)&=\hphantom{-}\sqrt{\mbox{$\frac{3}{2}$}}M\sqrt{1+\tau}
\left[
G_E(Q^2)P^{(M1,M1)0}(Q^2)-\sqrt{6}\,G_M(Q^2)P^{(L1,L1)1}(Q^2)
\right]\,,\label{eq:plt}
\end{align}
with $G_E(Q^2)$ and $G_M(Q^2)$ being the Sachs electric and magnetic form factors of the nucleon, and $\tau = Q^2/(4M^2)$.
Furthermore, if the electron polarisation transfer $\epsilon$
epsilon is kept constant, the cross section depends only on two combinations,
namely, $P_{LL}-P_{TT}/\epsilon$ and $P_{LT}$.
\begin{figure}[h]
\begin{minipage}{0.4\textwidth}
\centering
       \includegraphics[width=\textwidth]{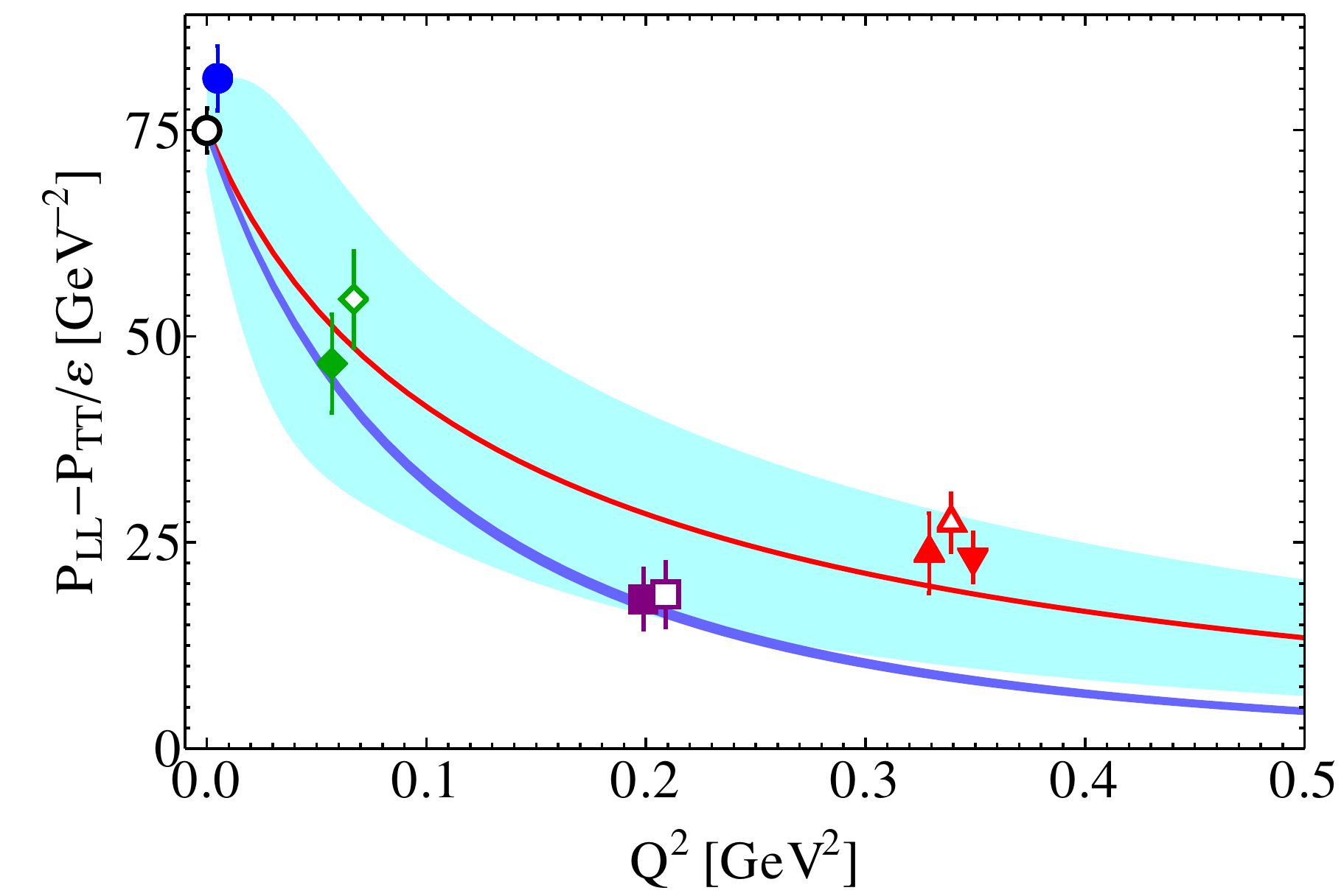}
    \end{minipage}\hfill
\begin{minipage}{0.4\textwidth}
\centering
  \includegraphics[width=\textwidth]{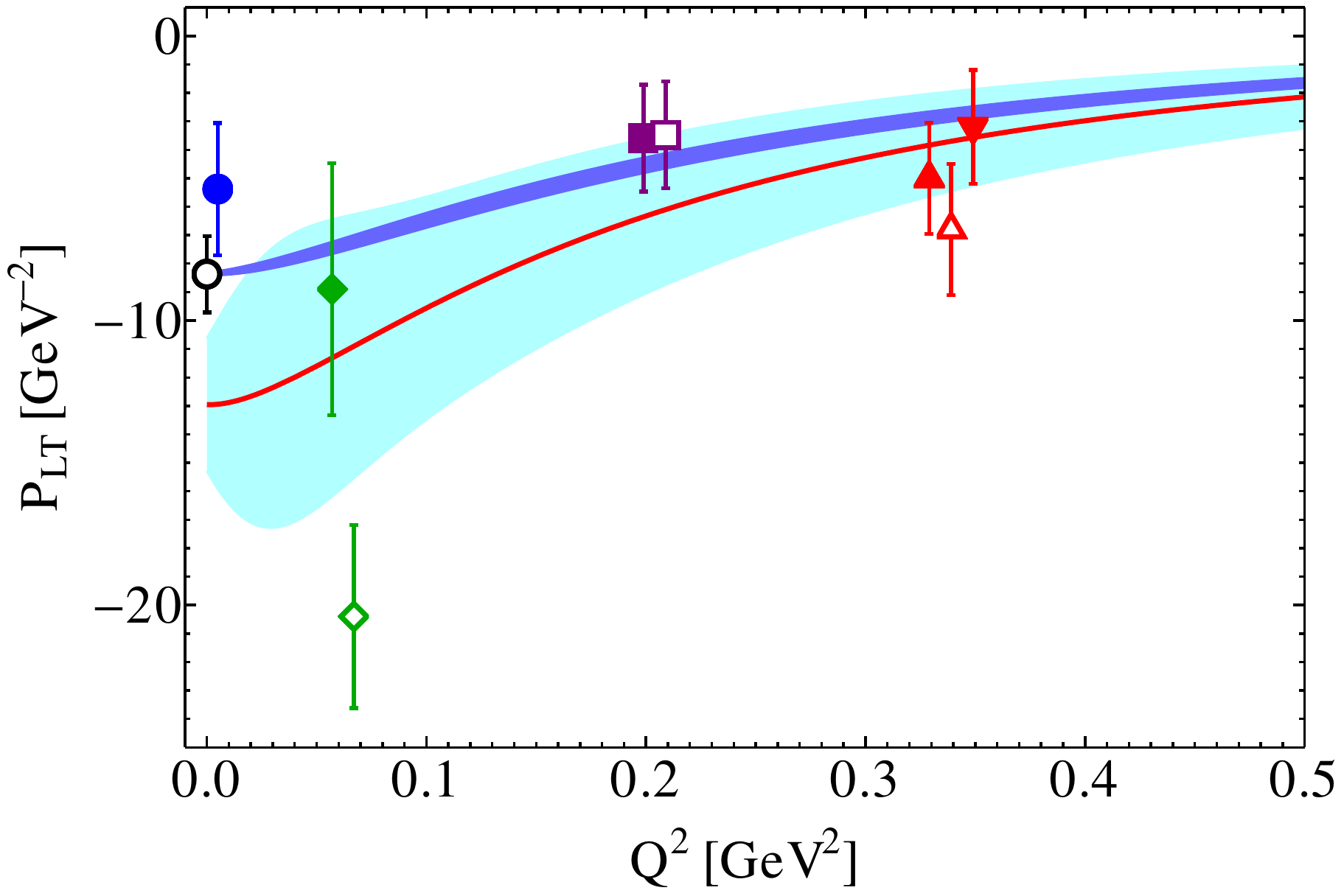}
\end{minipage}
\caption{VCS response functions $\displaystyle P_{LL}(Q^2)-P_{TT}(Q^2)/\varepsilon$ (left) and $\displaystyle P_{TT}$ (right). The curves correspond to $\varepsilon=0.65$.
The NNLO B$\chi$PT result --- red solid curve with the cyan band.
DR results~\cite{Drechsel:2002ar} --- blue band. 
The data shown are: real-photon values, black open circle, PDG 2014~\cite{Agashe:2014kda},
and blue circle, MAMI~\cite{MAMI01};
green diamond (solid/open), MIT-Bates~\cite{Bourgeois:2006js,Bourgeois:2011zz};
purple square (solid/open), MAMI~\cite{Correa:thesis};
red triangle (solid/solid inverted/open), MAMI~\cite{Roche:2000ng,d'Hose:2006xz,Janssens:2008qe};
see ref.~\cite{Lensky:2017dlc} for further explanation of the experimental points.}
\label{fig:gp_vcs}
\end{figure}

Figure~\ref{fig:gp_vcs} shows the comparison of the B$\chi$PT results for $P_{LL}-P_{TT}/\epsilon$ and $P_{LT}$ with the results of a DR calculation and the available
empirical extractions. The general agreement both with data and the DR calculation is quite good (within the somewhat large
theoretical errors). The tensions in $P_{TT}$ at low $Q^2$ are due to the difference in the static dipole polarisability $\beta_{M1}$.
To further test the $\chi$PT calculations in this sector, it would be very desirable to have more data at low $Q^2\lesssim 0.2~\text{GeV}^2$.
One of the interesting features to study there would be the slope of $\beta_{M1}(Q^2)$ at $Q^2 = 0$; this quantity enters the
spin-independent constraints below. The theoretical prediction for the slope is sensitive to the cancellation of the $\Delta$ pole contribution
with that of the $\pi N$ loops, so even the sign of this quantity is not well established. 

\section{Nucleon VVCS and generalised polarisabilities in B$\chi$PT}

In the case of (forward) VVCS, the amplitude is expressed in terms of only four scalar functions that depend on the photon 
virtuality $Q^2$ and the photon laboratory frame energy $\nu$~\cite{Drechsel:2002ar}:
\begin{align}
\hspace*{-1.em}T=&\;f_{L}(\nu,Q^2) +(\vec{\epsilon}^{\, \prime *} \cdot \vec{\epsilon}\,) \,f_{T}(\nu,Q^2) 
+
  \; i \vec{\sigma}\cdot (\vec{\epsilon}^{\, \prime *} \times \vec{\epsilon}\,)\, g_{TT}(\nu,Q^2) -  i \vec{\sigma}\cdot [(\vec{\epsilon}^{\, \prime *} - \vec{\epsilon}\,)\times \hat{q}]\, g_{LT}(\nu,Q^2)\,. 
\end{align}
The low-energy expansion (LEX) of the non-Born part of the scalar amplitudes (denoted with the bar over the corresponding letters) is 
\begin{align}
  \bar{f}_T(\nu, Q^2) &= 4\pi\left[ Q^2 \beta_{M1} +{(\alpha_{E1} + \beta_{M1})} \nu^2\right] + \dots,\quad 
  &\bar{f}_L(\nu, Q^2) &=  4 \pi(\alpha_{E1}+{\alpha_{L}} \nu^2) Q^2 + \dots \nonumber\\
 \bar{g}_{TT}(\nu, Q^2) &= 4 \pi{\gamma_0}  \nu^3 + \dots ,\quad
  &\bar{g}_{LT}(\nu, Q^2)&=  4 \pi{\delta_{LT}} \nu^2 Q + \dots ,
\end{align}
where the longitudinal polarisability $\alpha_L$ and the longitudinal-transverse polarisability $\delta_{LT}$
are new for the VVCS case, whereas the other polarisabilities are the same as in the RCS case, 
with $\gamma_0=-\gamma_{M1M1}-\gamma_{M1E2}-\gamma_{E1E1}-\gamma_{E1M2}$. These static polarisabilities can be naturally generalised
considering them as functions of $Q^2$, such as $4\pi Q^2 \beta_{M1}(Q^2) =\bar{f}_T(0,Q^2)$ and so on.
Again, these generalisations are specific to the regime of VVCS.
The $Q^2$-dependent coefficients in front of non-zero powers of $\nu$ in the LEX of the scalar amplitudes are related through the unitarity
with integrals of electroabsorption cross sections, or the nucleon structure functions. 
\begin{figure}[h]
\centering
 \begin{minipage}{0.8\textwidth}
\centering
       \includegraphics[width=\textwidth]{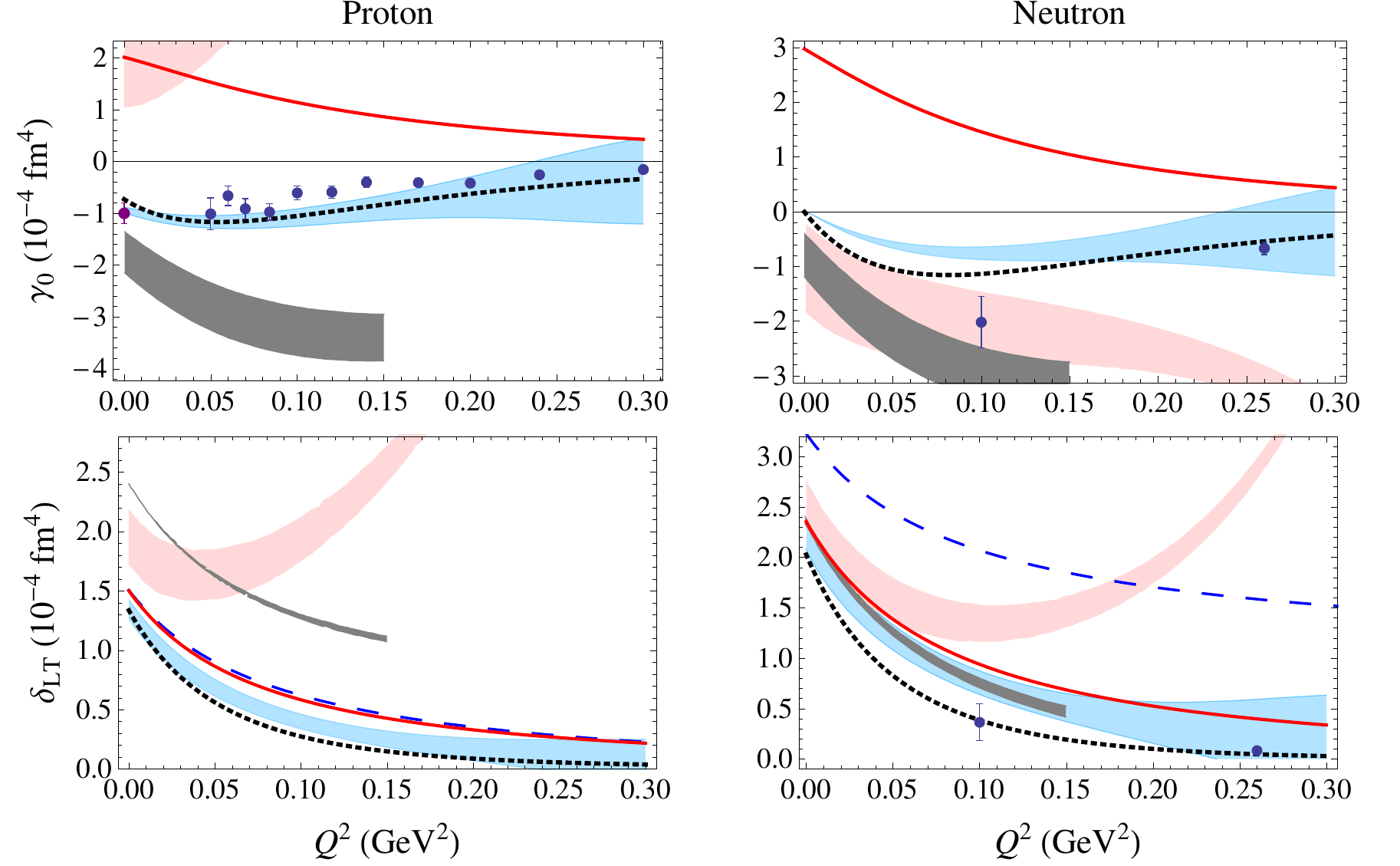}
       \end{minipage}
\caption{Generalised polarisabilities $\gamma_0(Q^2)$ and $\delta_{LT}(Q^2)$ of proton and neutron.
Red solid lines and blue bands: NLO and NNLO B$\chi$PT result. Black dotted lines: MAID2007~\cite{Drechsel:2007if}.
Grey bands: B$\chi$PT  calculation of ref.~\cite{Bernard:2012hb}. Blue dashed line: NNLO HB$\chi$PT calculation~\cite{Kao:2002cp}; off the scale for $\gamma_0(Q^2)$.
Red band: infrared-regularised B$\chi$PT calculation~\cite{Bernard:2002pw}. The data points for the proton $\gamma_0$ at finite $Q^2$ are from Ref.~\cite{Prok:2008ev} (blue dots), and at $Q^2=0$ from~\cite{Dutz:2003mm} (purple square).
For the neutron all the data are from Ref.~\cite{Amarian:2004yf}. \label{Fig:GridSpin}}
\end{figure}

The spin VVCS GPs such as $\gamma_0(Q^2)$ and $\delta_{LT}(Q^2)$ have been extensively mapped within the JLab low-$Q^2$ experimental programme,
reviewed in refs.~\cite{Deur:2018roz,Deur_talk}. Of these two, $\delta_{LT}$ is assumed to receive only a tiny
contribution from the $\Delta$, and therefore is an excellent check of chiral dynamics.
The first HB$\chi$PT calculations, however, could not describe the early data on the neutron $\delta_{LT}$~\cite{Deur_talk} --- the ``$\delta_{LT}$ puzzle''.
Figure~\ref{Fig:GridSpin} shows the current status of $\gamma_0$ and $\delta_{LT}$.
One can see that the B$\chi$PT results of ref.~\cite{Lensky:2014efa} provide a much better description
of the available information on the proton $\gamma_0(Q^2)$, compared with the HB$\chi$PT and the infrared-regularised
B$\chi$PT. At the same time, preliminary data on the neutron/deuteron show a disagreement
especially in the region of low $Q^2$~\cite{Deur_talk,Adhikari:2017wox}; one of the possible explanations
suggested recently is the missing few-nucleon contributions in the analysis of the deuteron data~\cite{Lensky:2018vdq}.
One also has to note that there is a different B$\chi$PT calculation~\cite{Bernard:2012hb}
of the VVCS GPs that uses a different counting and, as a result, has a different set of $\pi\Delta$ loops
(and also treats the $\Delta$ pole contribution differently, most notably, does not include
the form factors). The two B$\chi$PT calculations disagree quite substantially, especially
in the values of $\delta_{LT}$. From the theory side, this discrepancy will most likely be resolved
in a higher-order calculation; at the same time, the new experimental data from JLab~\cite{Deur_talk} will
test the theory.

The scalar VVCS amplitudes and GPs are interesting, first of all, due to their contribution in the Lamb shift (LS) of muonic hydrogen ($\mu$H),
see, e.g.~\cite{Hagelstein:2015egb} for a review. The dispersion relation for $f_{T}(\nu,Q^2)$ that relates
it to the nucleon structure functions involves a subtraction function $\bar{T}_1(Q^2)=\bar{f}(0,Q^2)$,
which cannot be calculated from a sum rule.
This function is readily obtainable from the B$\chi$PT calculation of the nucleon CS,
with the fact that the B$\chi$PT results are consistent with the wealth of available data on the nucleon CS
providing an additional verification of the result.
The B$\chi$PT result for the polarisability contribution to the $\mu$H LS is $-8.2_{-2.5}^{+1.2}$~$\mu$eV at NLO ($\pi N$ loops~\cite{Alarcon:2013cba})
which shifts to $-7.3^{+1.5}_{-2.7}$~$\mu$eV when the $\Delta$ pole contribution is added~\cite{Lensky:2017bwi},
which agrees well with the results of other approaches.

\section{Analyticity constraints relating RCS, VCS, and VVCS}

As mentioned in the introduction, the analyticity of the most general (non-Born, i.e., with the Born part subtracted) CS amplitude
means that the LEX of this amplitude is universal: all kinematics, in particular, RCS, VCS, and VVCS,
are described by a single set of coefficients. This leads to constraints that relate polarisabilities between the different regimes. Several new constraints of that kind have been derived recently~\cite{Pascalutsa:2014zna,Lensky:2017bwi,Lensky:2017dlc}.
We start from two constraints
that connect spin-dependent quantities and appear at the order $k^2$ in the expansion in powers of small momenta $k$:
\begin{align}
I_1^\prime(0)& =  \frac{ \kappa_N^2}{12} \langle r_2^2 \rangle 
+ \frac{M^2}{2} \left\{ \frac{4\pi}{e^2} \gamma_{E1 M2} 
- 3 M \left[ P^{\prime \, (M1, M1)1}(0) +  P^{\prime \, (L1, L1)1} (0) \right]  \right\},
\label{eq:sr1}
\\
\delta_{LT} &= 
- \gamma_{E1 E1} 
+ \frac{3 M e^2}{4\pi} \, 
\left[ P^{\prime \, (M1, M1)1}(0) -  P^{\prime \, (L1, L1)1} (0) \right],
\label{eq:sr2}
\end{align}
where $I_1'(0)$ is the slope of the the generalised GDH integral~\cite{Drechsel:2002ar} at $Q^2=0$, $\kappa_N$ is the 
anomalous magnetic moment of the nucleon, $r_2$ is its Pauli radius, whereas  $P^{\prime \, (M1, M1)1}(0)$
and $P^{\prime \, (L1, L1)1} (0)$ are the slopes of the VCS GPs; note that $\delta_{LT}$ here also stands for the static value, $\delta_{LT}(0)$.
These two constraints are sum rules:
they relate measurable quantities --- moments of the nucleon structure functions --- with linear combinations
of polarisabilities.
\begin{figure}[h]
\begin{minipage}{0.4\textwidth}
\centering
       \includegraphics[width=\textwidth]{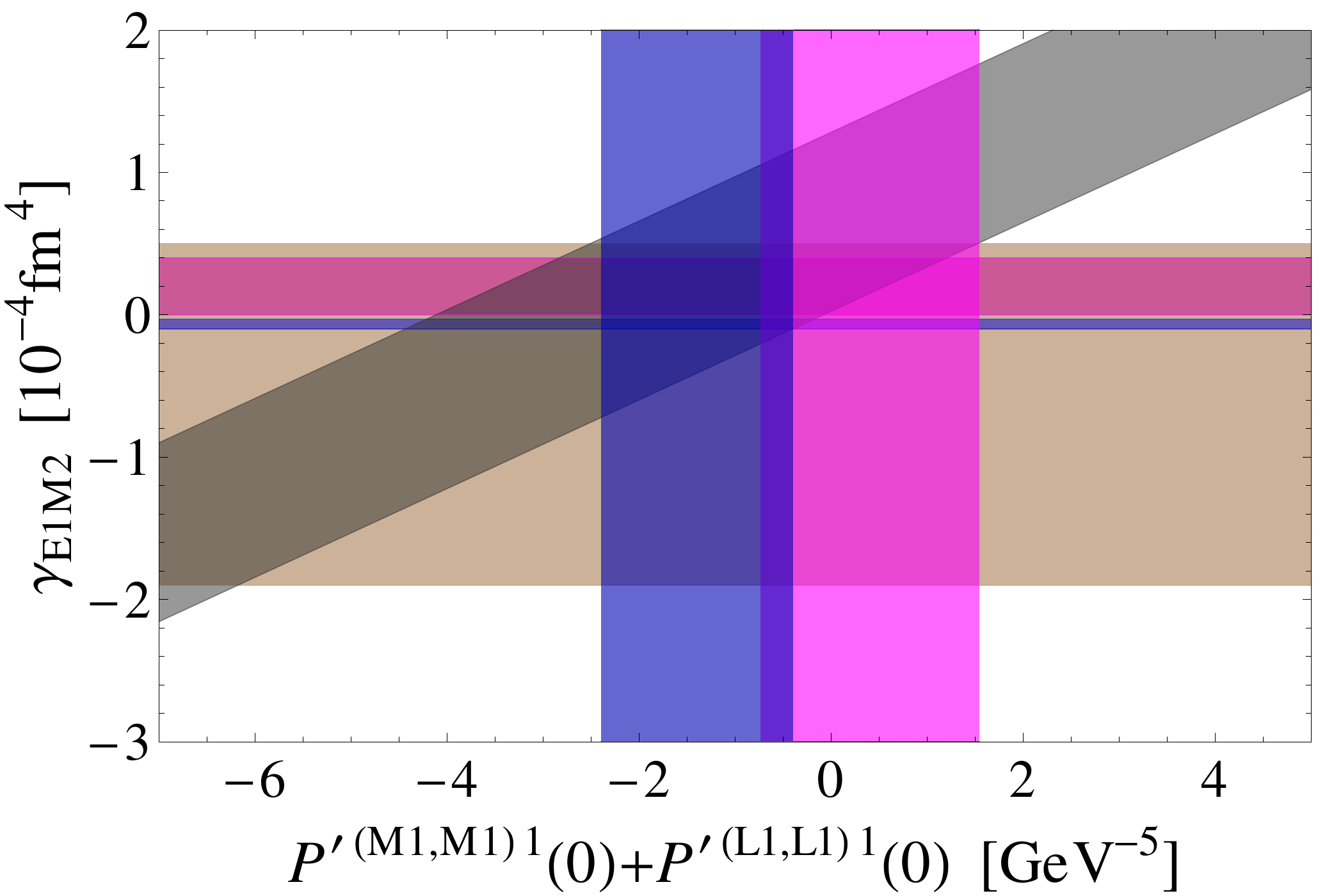}
    \end{minipage}\hfill
\begin{minipage}{0.4\textwidth}
\centering
  \includegraphics[width=\textwidth]{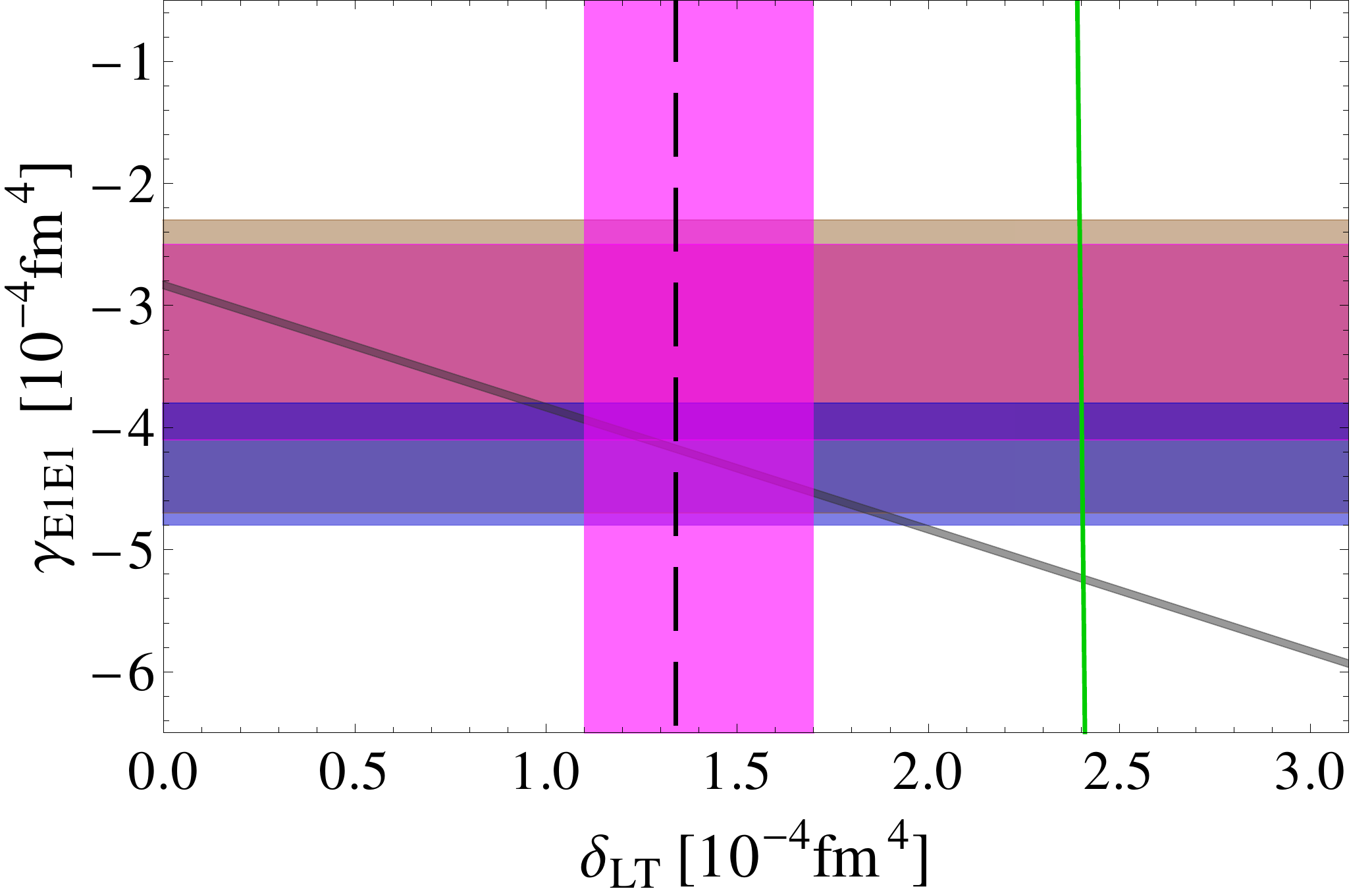}
\end{minipage}
\caption{Graphical representation of the spin-dependent sum rules of Eq.~\ref{eq:sr1} (left) and~\ref{eq:sr2} (right).
The brown bands in both panels are the empirical extraction of $\gamma_{E1 M2}$ and $\gamma_{E1E1}$~\cite{Martel:2014pba}. 
The blue bands are the DR evaluations~\cite{Drechsel:2002ar} for the RCS and the slopes of VCS polarisabilities.
The magenta bands are the B$\chi$PT evaluations~\cite{Lensky:2014dda,Lensky:2015awa,Lensky:2016nui}.
The gray band in the left panel is the sum rule constraint based on the empirical information for $I_1^\prime(0)$ and $ \langle r_2^2 \rangle$.
The gray band in the right panel is the sum rule constraint based on the DR evaluation for the slopes of the VCS polarisabilities,
whereas the black dashed line and the blue solid line are respectively the values of $\delta_{LT}$ resulting from the analysis of 
MAID2007~\cite{Drechsel:2007if} and the B$\chi$PT calculation of ref.~\cite{Bernard:2012hb}. See ref.~\cite{Lensky:2017dlc} for more detail.}
\label{fig:spin_sum_rules}
\end{figure}
Figure~\ref{fig:spin_sum_rules} shows a graphical representation of these two constraints. In particular,
the right panel of that figure again illustrates the $\delta_{LT}$ puzzle, where the B$\chi$PT result of ref.~\cite{Bernard:2012hb}
shows some tension with the other evaluations.

Three further new constraints arise at the order $k^4$ in the expansion of the scalar CS amplitudes. They take the form~\cite{Lensky:2017bwi}
\bea
M^{ (2)\prime}_1(0)  &=& 
 \beta_{M1, \nu} + \frac{1}{12} (4 \beta_{M2} + \alpha_{E2}) 
 +2 
\left( \alpha_{E1}^\prime +\beta_{M1}^\prime \right)
- \frac{e^2}{2\pi} (2M)^2  b_{4, 1}
  \nonumber \\
&+& \frac{1}{M} \left(- \delta_{LT} + \gamma_{M1M1} - \gamma_{E1E1} - \gamma_{M1E2} +  \gamma_{E1M2} \right) 
+ \frac{1}{(2 M)^2} (\alpha_{E1} + \beta_{M1}), \quad \quad
\label{eq:mp12}\\
M^{(1)\prime}_2(0) &=&  \frac{1}{6} (\alpha_{E2} + \beta_{M2}) 
+2 
\left( \alpha_{E1}^\prime +\beta_{M1}^\prime \right)
-  \frac{e^2}{4\pi} \,(2 M)^2 b_{19, 0}
\nonumber \\
&&- \frac{1}{M} \left( \delta_{LT}  + \gamma_{E1E1} + \gamma_{M1E2} \right)  
+ \frac{1}{(2 M)^2} (\alpha_{E1} + \beta_{M1}),
\label{eq:mp21}\\
\frac{1}{2}\bar{T}_1^{\prime\prime}(0)
&=&\frac{1}{6} \beta_{M2} 
+2\beta_{M1}^\prime  
 + \frac{e^2}{4\pi} b_{3, 0}
 + \frac{1}{(2 M)^2} \beta_{M1}  \,,
\label{eq:T1pp}
\eea
where $M_1^{(2)\prime}(0)$ and $M_2^{(1)\prime}(0)$ are the slopes of the second moment of the unpolarised structure function $F_1(x,Q^2)$ and of the first moment
of the unpolarised structure function $F_2(x,Q^2)$, $\bar{T}_1^{\prime\prime}(0)$ is the second derivative of the VVCS subtraction function,
and $\alpha'_{E1}$, $\beta'_{M1}$ are the slopes of the generalised VCS polarisabilities at $Q^2=0$. The first two of these constraints
are sum rules; 
the first of them involves a new constant $b_{4,1}$ which is accessible in VCS through higher-order GPs, while the second sum rule
contains the constant $b_{19,0}$ which, in principle, could be accessed in an off-forward VVCS process (such as, e.g., lepton pair
electroproduction). The third constraint involves a derivative of the subtraction function, rather than a moment of a structure function,
and therefore cannot be represented as a sum rule. The coefficient $b_{3,0}$ that enters it can also be accessed in an off-forward VVCS process,
and its knowledge (together with that of the other polarisabilities in the right-hand side of the constraint)
would allow one to empirically constrain the slope of the subtraction function, see the discussion in ref.~\cite{Lensky:2017bwi}.

\section{Summary}

We reviewed the current status of the covariant B$\chi$PT results for the different regimes of nucleon CS
and the polarisabilities that are characteristic of these regimes. The calculations done in the same framework
allow one to test the theory by comparison with empirical data in all these sectors. This, in turn, verifies the B$\chi$PT calculations
of the inelastic structure contributions to the two-photon exchange corrections, in particular, to
the Lamb shift of muonic hydrogen. We showed that the agreement between the B$\chi$PT results
and empirically available data is rather good generally, especially in RCS. There are still puzzles
such as the difference between the extracted values of $\beta_{M1}$ in the different analyses, and the description
of the VVCS spin GPs (the $\delta_{LT}$ puzzle) where there is a significant variation between different
theory calculations. While the disagreement between the theory results will likely be resolved in future higher-order B$\chi$PT calculations,
new experimental data expected from different facilities (MAMI, JLab, HIGS) will further test
the $\chi$PT description. Additionally, we reviewed new model-independent constraints that relate the polarisabilities across the different regimes 
of nucleon CS; these constraints could provide additional information on the low-energy structure of the nucleon.

\end{document}